\documentstyle[12pt]{article}
\title{\Large  F-theory  description  of  3-string  junction }
\author{\large  Fu-Zhong Yang \\
\small Department of Physics, The Graduate School of the Chinese Academy of
Sciences,
\\ \small  PO Box 3908, Beijing, 100039, The People's Republic of China}
\date{}
\def\ZZ{{\sf Z\!\!Z}}
\begin{document}
\maketitle
%\noindent
\vspace{1cm}
\indent
\begin{center}
  \begin{minipage}{4.8in}
\begin{center}
 {\large\bf Abstract}
\end{center}
\vspace{1cm}
%\baselineskip 0.8cm
%\normalsize
\small  The geometrical description of BPS $3$-string junction in
the F-theory background is given by lifting a string junction in
IIB into F-theory and constructing a holomorphic curve in $K3$
with respect to a special complex structure of K3. The holomorphic
curve is fibration of 1-cycles of the elliptic fiber over the
geodesic string junction. The F-theory picture in this paper
provides an unifying description of both string and string
junction, and is advantageous over the M-theory picture of them.
  \end{minipage}
\end{center}
%-----------------------------------------------
\newpage
\section{\large Introduction}

Over the last years, our understanding of the non-perturbative
features of supersymmetric field and string theories has undergone
a dramatic change. The key to this development is the discovery of
duality symmetries which connect at first sight seemingly
completely unrelated two theories. The central role in this
impressive progress has been played by BPS staturated soliton
objects in the spectrum of the theory. In particular, the D-brane
has proven to be a powerful non-perturbative tool, and provided a
geometrical description of states in field and string theories.
When $N$ D-branes coincide, we get enhanced gauge symmetry $U(N)$
in the world-volume field theory of the branes , the gauge vectors
of which are open strings stretched between these D-branes in all
possible ways. The charges carried by an open string are
determined by the branes it ends on. When a pair of D-branes are
separated, the open string stretched between them gets mass
proportional to the separation of the two D-branes. This is just
the D-brane picture of the standard Higgs mechanism.

However, open string description of the gauge vector of
exceptional gauge group does not seem to be suffice, the
multi-pronged open strings with more than two endpoints, in
particular, the three-string junctions, are necessary.  Moreover,
the multi-pronged open strings states manifestly carry the correct
gauge charges. This case arise in the compactification of IIB
superstring on a two-sphere in the presence of mutually non-local
$7$-branes, not all of which are of the same [p,q] type. In fact,
when an ordinary string crosses the 7-brane it can't end on, a
one-brane is created. In this process, a $3$-string junction can
arise \cite{1}-\cite{6}. In fact, the string creation phenomenon
is completely analogous to the Hanany-Witten effect \cite{7}.
Thus, the string creation mechanism give rise to a transition from
string to string junction. Thus, it has become quite obvious that
in type IIB superstring theory, (p,q) string and string junction
are actually on the same footing. Of interest is the fact that the
3-string junction preserves one fourth of the supersymmetries, and
is thus stable. Geometrically, the BPS multi-pronged open string
junction is just the geodestic curves meeting at one point on the
sphere.

On the other hand, the compactification of the IIB superstring on
a two-sphere in the presence of twenty-four mutually non-local
7-branes is dual to the F-theory on $K3$ surface fibered
elliptically on the sphere while the complex structure of the
elliptic fiber ( torus $T^{2}$ ) corresponds to the IIB complex
coupling, which vary as one move on the base.  The points on the
base  where the complex structure become singular represent the
positions of the [p,q] $7$-branes in IIB.

Twenty-four such points are required in order to preserve some
supersymmetry and further to get a smooth $K3$. It will become
singular only when some of the points coalesce. When this happens,
some non-contraclible $2$-cycles of K3 shrink to zero size. In
general, such $2$-cycle is a fibration of a (p,q) $1$-cycle of the
torus over a line segment connecting two points associated to the
positions of two [p,q] $7$-branes on the base. Therefore, from the
viewpoint of F-theory, a (p,q)-open string stratched between two
[p,q] $7$-branes and multi-pronged string junction are nothing but
the 2-cycles in the K3 respectively. Furthermore, if the string or
junction present BPS state, the associated 2-cycle is holomorphic
curve in the elliptically fibered K3. This kind of the 2-cycles is
the F-theory description of BPS string or BPS string junction in
IIB.

String junction is originally introduced in \cite{8,9}, the BPS
nature is studied in \cite{5,6,10,11,12,22}. Moreover, their
dynamics and application to SUSY quantum field theory are studied
in \cite{21} and \cite{23}-\cite{29}. These properties are studied 
from M(atrix) theory viewpoint in \cite{39}. The algebrac classification
of them with support on various 7-brane backgrounds is extensively
investigated in \cite{10}-\cite{21}. However, the Ref.\cite{21}
etc. did not explore the M-theory or F-theory picture of the
string junction.

The geometrical properties of junctions from M-theory viewpoint
were given in \cite{30,31,32}, which have discussed infinitely
stretched BPS string configuration in the 7-brane backgrounds, but
have not studied finitely stretched BPS string junction. However,
as pointed out in \cite{32}, for further understanding of the BPS
states and the 3-string junction generating process in the 7-brane
backgrounds, an M-theory description of finitely stretched BPS
string configurations ending on 7-branes would be desired. But,
strictly speaking, the M-theory description cannot be trusted when
the compactifiction tours is smaller than the 11-dimension Planck
scale \cite{30}. Thus, it is meaningful to explore the junction
from the F-theory viewpoint and the geometrical engineering
program.

In this short note, we shall give a F-theory description of
finitely stretched BPS 3-string junction. Concretely, we lift a
string junction in IIB into F-theory, and construct a holomorphic
curve in K3 with respect to a special complex structure of K3. The
holomorphic curve is fibration of 1-cycles of the elliptic fiber
over the geodesic string junction  which consist of finitely
stretched strings on the base $CP^1$  of the K3 surface. Thus, the
F-theory description of the junction is advantageous over the
M-theory picture.

The rest of the paper is organized as follows. In Section 2,  we
briefly review some of the relevant features on F-theory and
junction that we will need. In Section 3, we construct the
F-theory description of the $(p,q)$-string and 3-string junction.
In Section 4, we give a short discussion.

\section{ F-Theory, Dualities and Junction }
\subsection{Review on the junction }

In the massless bosonic sector of IIB superstring, there are the dilaton $\phi$ and
the axion $\chi$ with their combination being the complex coupling:
\begin{equation}
 \tau = \tau_{1} + i\tau_{2} = \chi + i e^{-\phi}
    \label{no1}
\end{equation}
The non-perturbative $SL(2, \ZZ)$ symmetry of IIB string acts on it in the canonical way :
\begin{equation}
\tau \rightarrow \frac{a \tau + b}{c \tau + d}; ~~~~~~~
 a, b, c, d \in {\large \ZZ}, ~ ad-bc = 1
\label{no2}
\end{equation}
Due to this symmetry, In the spectrum of IIB superstring, there are
a set of various (p,q) strings, here p and q are relatively prime integers denoting
the charges of the string under NS-NS and RR antisymmetric tensors respectively.
A D7-brane carries one unit of D7-brane charge, it is just a unit source of 8-form tensor
field dual to the axion $\chi$. This means that it is magnetically charged with respect to
the the axion $\chi$. A [p,q] seven-brane is a seven-brane where a (p,q) string can end.
The tension of a (p,q)-string is :
\begin{equation}
          T_{p,q} = \frac{1}{\sqrt{\tau_{2}}} | p + q\tau | ,
  \label{no3}
\end{equation}
We associate a branch cut (on the sphere) and a monodromy $M_{p,q}$ to
the [p,q] seven-brane in the transverse direction. When an (r,s)-string crosses
this cut in a counter-clockwise way it undergoes the following monodromy transformation:
\small
\begin{equation}
           \left( \begin{array}{c}r\\s \end{array}  \right) \rightarrow
   M_{p,q} \left( \begin{array}{c}r\\s \end{array}  \right)=
       \left( \begin{array}{c}r\\s \end{array}  \right) +
(sp-qr)\left( \begin{array}{c}p\\q \end{array}  \right),~~~
         M_{p,q} = \left( \begin{array}{cc}1-pq & p^{2}\\
                                          -q^{2}& 1+pq
\end{array}
\right)
\label{no4}
\end{equation}
\normalsize
In this process, $|sp-rq|$ (p,q)-strings are created between the [p,q] seven brane and
this string, which are oriented so as to conserve the charges. When $|sp-rq|=1$,
a 3-string junction arises.

A 3-string junction is a configuration where three strings with their charges
being $(p_{i},q_{i})$ meet at a point, and satisfies the charge conservation:
$$           \sum^{3}_{1} p_{i}= \sum^{3}_{1}q_{i} = 0    $$

These conditions ensure the equilibrium of the force (due to the
tensions of strings) at the vertex.

Gauge vector is realized sometimes as geodesic string, but more
generally, it is realized as a geodesic string junction. This is
the IIB picture of the string junction. On the other hand, IIB
string can be viewed as M-theory on a torus in the limit where the
torus shrinks. The string is the membrane of M-theory with one
direction wrapped on the torus. The (p,q)-string is determined by

the homology cycle of the torus the membrane wraps. Thus, there
are a  M-theory picture of string junction \cite{30,31,32}.
However, strictly speaking, the membrane description cannot be
trusted when the compactifiction tours is smaller than the
11-dimension Planck scale \cite{30}. Thus, it is also meaningful
to explore the junction from the F-theory viewpoint.

\subsection{ F-theory and Duality }

F-theory is another novel way of generating non-perturbative  compactification.
This compactification is intrinsically non-perturbative in the sense that
the string theory coupling necessarily becomes strong in some regions of
the moduli space of compact manifold, unlike conventional compactification
where the string coupling can be kept small everywhere on the internal manifold.

One of the motivations for F-theory\cite{29,33,34,35} is to find a far more satisfactory
understanding of $SL(2,\ZZ)$ symmetry of IIB string in ten dimensions, and to get
the geometric meaning directly in ten dimensions.

From M-theory viewpoint, if we go to nine dimensions, we can see
the $SL(2,\ZZ)$ symmetry as a geometric symmetry of a torus, ---
the modular group. However, in order to go to the ten dimensional
limit, we have to take the limit of zero size torus, which is not
natural, and gets us away from the domain of validity of geometric
description of M-theory. Thus it is natural to postulate the
existence of 12D theory ---- F-theory,
 which was found by Vafa in constructing the compact D-manifold for type IIB string in 1996. He found new non-perturbative vacuum of IIB string. More precisely,
 as the first Chern class of the 2-sphere $CP^{1}$ does not vanish,
 this space can not be a good solution, in particular, it does not preserve any SUSY.
 To get a consistent background, we can put extra 7-branes on the theory,
 which sit at arbitrary points $z_{i}$ on the $CP^{1}$, and fill the 7+1 non-compact dimensions.
 As a consequence, when looping the location $z_{i}$, there will be
 a highly non-trivial monodromy in the $SL(2,\ZZ)$. Thus it is natural to regard
 the complex coupling as modular parameter of a two-torus $T^{2}$, furthermore,
 the modular group of the torus as the $SL(2,\ZZ)$ symmetry of IIB.
 The relevant geometrical picture is just a fibration of torus over $CP^{1}$.

From the Kodaira's formula for the canonical bundle of the total space $S$ which is
a elliptic fibration over the base $C$, we will get the right way.
The Kodaira's formula[36] is as follows:
\begin{equation}
            K_{S} = \pi^{*} \left(K_{C} + \sum a_{i} P_{i}\right) \label{no5}
\end{equation}
where $K_{S}$ and $K_{C}$ are the canonical bundles of the totale
space $S$ and base $C$, respectively. The sum is over points
$P_{i}$ of the base $C$ of the elliptic fibration $S\rightarrow
C$, the fibers are singular at the points $P_{i}$. The
coefficients $a_{i}$ are determined by the properties of the
singular fiber. For the case in considering, we take the base
$C=CP^{1}$ with $C_{1}(K_{C}) = -2$ , and take the coefficients
$a_{i} = 1/12$, then we must put $24$ points marked on the base
$CP^{1}$ in order to have a consistent IIB background with
$C_{1}(K_{S})=0$. This is just a smooth K3 surface. Thus,
compactification of the IIB superstring on a two-sphere in the
presence of twenty-four mutually non-local 7-branes is dual to the
F-theory on K3  surface fibered elliptically on the sphere while
the complex structure of the elliptic fiber ( torus $T^{2}$ )
corresponds to the IIB complex coupling, which vary as one move on
the base.

In the Weierstrass form, the elliptic K3 can be represented as:
\begin{equation}
       y^{2} = x^{3} - f(z) x - g(z)   \label{no6}
\end{equation}
where $z$ is the coordinate on the sphere, in general, the functions $f(z)$ and $g(z)$ are
polynomials of degree 8 and 12 in $z$ respectively. The $J$-invariant of the elliptic fiber
at the point $z$ is as follows:
\begin{equation}
J(\tau(z)) = \frac{ 4(24 f(z))^{3}}{ (4 f^{3}  -  27 g^{2})}  \label{no7}
\end{equation}
the $J$-invariant is defined as:
\begin{equation}
J(\tau) = \frac{(\theta^{8}_{1} + \theta^{8}_{2} + \theta^{8}_{3})^{3}}
{\eta(\tau)^{24}}
            \label{no8}
\end{equation}
where $\theta_{i}$ is the theta function, and $\eta(\tau)$ is the Dedekind eta-function:
\begin{equation}
\eta(\tau) = q^{1/24} \prod^{\infty}_{n=1} (1 - q^{n}),
q = exp(i2\pi \tau)   \label{no9}
\end{equation}
In general, the $\tau (z)$ can be derived from equations \ref{no7}-\ref{no8}
though it is not easy. The discriminant of the equation \ref{no6} is
just the denominator of the fraction in the equation  \ref{no7} as follows:
\begin{equation}
\bigtriangleup =  4 f^{3}  -  27 g^{2}       \label{no10}
\end{equation}
The points where $\bigtriangleup = 0$ are the positions of the singular fibers
with $J(\tau) \rightarrow \infty$, and correspond to the locations of 7-branes.
By the way, the number of the independent variables in equation  \ref{no10} is
only 18, not 24. Thus, the positions of the 24 7-branes cann't be all independent,
but must be restricted topologically. In fact, there are a global constraint
on the monodromies of the 24 7-branes from the fact that a loop around
the entire configuration of 7-branes on the sphere can shrunk to
a point on the other side of the sphere. As a consequence,
the product of monodromies of all the 7-branes must be trivial.
Furthermore, the topological consistency turns out to support only at most 18
mutually commuting monodromies, thus at most 18 independent $U(1)$ factors
in the symmetrical group of the full theory.

If two 7-branes collide, both the 2-cycle of K3 and open string
between them on the base obviously shrink to zero size. This lead
precisely to the geometrical singularity of K3 surface on the side
of the F-theory while we get enhanced gauge symmetry from D-brane
picture of the standard Higgs mechanism. This is just an
indication of the so called geometric engineering which uncover
the correspondence between the geometric singularity and the
enhanced gauge symmetry. In particular, the Kodaira singularity of
the elliptic fiber of K3 is exactly associated to enhanced gauge
symmetry on the side of F-theory. Contrarily, separating the
coincided 7-brane is associated to resolving the singularity by
blown up mathematically and breaking gauge symmetry physically.
Thus, blown up can be regarded as the geometric realization (or
F-theory realization) of standard Higgs mechanism. In this
process, the strings and/or string junctions stretched among these
7-brane get masses.

F-theory has led to many insights into string dualities in an elegant way.
One of the most basic dualities is the following one:
F-theory on K3 is dual to the heterotic string on the torus $T^{2}$, more generally,
 F-theory on a Calabi-Yau $(n+1)$-fold $W_{n+1}$ is dual to the heterotic string
 on a Calabi-Yau $n$-fold $Z_{n}$ , where $K3\rightarrow W_{n+1}\rightarrow B_{n-1}$  is
 a fibration with the K3 being elliptic fibration,
 while $T^{2}\rightarrow Z_{n}\rightarrow B_{n-1}$ is also elliptic fibration.
    In F-theory, the string junction is just the 2-cycle of K3, more precisely,
    a fibration of 1-cycle of K3 over the junction on the base.
    In the next section, we will construct the F-theory picture of BPS junction.

\section{ F-theory description of 3-string junction }
\subsection{ F-theory description of (p,q)-string}

The M-theory description of (p,q)-string and 3-string junction were given
in Ref.\ \cite{30,31}. They consider M-theory on $R^{1,8}\times T^{2}$.

The M-theory description of (p,q)-string and 3-string junction
in 7-branes background were given in Ref.\cite{32}
by choosing a appropriate complex structure of the background geometry.
In this section, we will lift the M-theory description to F-theory.

As the coordinates of the elliptic fiber of the elliptic K3 are
essentially represented by one complex varible, the generic
complex structure of K3 is not appropriate to describe a string or
junction by holomorphic curve of K3 with the holomorphic 2-cycle
being fibration over the string or junction. We are now in the
position to transform the complex structure of K3
 so as to describe the string and junction as a two-cycle of K3.

Though F-theory involve a $(10,2)$ spacetime, in compactifying $(1,1)$ spacetime,
the physical degree of freedom is only the complex structure of the torus.
Thus, we can view it geometrically as if we are compactifying it on a Euclidean tours $T^{2}$.
 In fact, we can translate the geometry to that of a $(10,2)$ manifold via the map
 given in \cite{33}

Suppose that the torus $T^{2}= (X^{10}-X^{11}~ plane)$ modulo a lattice,
which is determined by a modular parameter $\tau$ and $X^{10}\sim X^{10} + 2\pi R$.
 Let $u= X^{10}+iX^{11}$. Just as before, $z$ is the coordinate on the sphere,
 the base of the K3. On the sphere, there is a $SL(2, \ZZ)$ invariant metric as follows:
\begin{equation}
  ds^{2} = \tau_{2} |w(z)dz|^{2};~~~~~~~ ~~~~~
w(z) = \tau(z)^{2}\prod^{24}_{i=1}(z-z_{i})^{-1/12}  \label{no11}
\end{equation}
 we make the following transformation of the complex structure
\begin{eqnarray}
U = Im \int^{z} \tau w(z)dz + i R^{-1}X^{11}   \label{no12} \\
V = - Im \int^{z} w(z)dz + i R^{-1}X^{10}     \label{no13}
\end{eqnarray}
Furthermore, make a transformation of the variables for convenience:
\begin{equation}
s = exp(RU) ~~~~~ \mbox{   and   } ~~~~~ t = exp(RV)     \label{no14}
\end{equation}
the equation of a (p,q)-string in the target space $S^{2}\times T^{2}$ can be written down:
\begin{equation}
        s^{q} t^{-p} = constant         \label{no15}
\end{equation}
So far, the above formulas are the analogy of Ref.\cite{30,32}. Now we will make another
 transformation of the Eqs.\ref{no12}-\ref{no13} in order to write down
 a equation of the holomorphic 2-cycle of K3, representing a (p,q)-string in IIB.
 Because the right side of Eqs.\ref{no12} -\ref{no13} only involve, in fact,
 two complex variables $u$ and $z$ , we can, in principle, solve out them as follows:
\begin{eqnarray}
              u = h (U,V) = H(s)            \label{no16} \\
z = p (U,V) = P(s)            \label{no17}
\end{eqnarray}
where the unknown functions $h$ and $p$ can be obtained from Eqs.\ref{no12}- \ref{no13},
furthermore, $H$ and $P$ can be got by substituting equation \ref{no14} and \ref{no15}
into function $h$ and $p$, respectively.

 Now we give the F-theory picture of a (p,q)-string by using
 the elliptic curve theory \cite{37,38} and embedding Eq.\ref{no15} into
 the elliptic K3 in Eq.\ref{no6} as follows:
\begin{eqnarray}
x & = & \frac{1}{(2\pi i)^{2}} \wp(u)
= \frac{1}{(2\pi i)^{2}} \wp(H(s))      \label{no18}  \\
y & = & \frac{1}{2(2\pi i)^{2}} \wp^{\prime}(u)
        = \frac{1}{2(2\pi i)^{2}} \wp^{\prime}(H(s))  \label{no19} \\
z & =  & P(s)        \label{no20}
 \end{eqnarray}
where the $\wp(u)$ and $\wp^{\prime}(u)$ are the famous Weierstrass $\wp$-function and
its derivative, respectively. The coordinates $x$ and $y$ satisfy the following equation:
 \begin{equation}
y^{2} =  x^{3}  - \frac{ g_{2}(\tau) }{4(2\pi i)^{4}} x -
\frac{g_{3}(\tau)}{4(2\pi i)^{6}}      \label{no21}
\end{equation}
where $ g_{2}(\tau) = \frac{4\pi^{4}}{3}E_{4}(\tau)$ and $
g_{3}(\tau) = \frac{8\pi^{6}}{27}E_{6}(\tau)$   with the
$E_{k}(\tau)$ being the standard normalized Eisenstein Series
\cite{37,38}. If we replace the complex structure $\tau$ by the
complex coupling $\tau(z)$ of the IIB string, then Eqs.\ref{no21}
is just the Eq.\ref{no6}. Thus, the Eqs.\ref{no18}-\ref{no21}
represent a holomophic curve embedded in the elliptic K3 with
respect to a special complex structure (Eqs.\ref{no12}-\ref{no14})
of K3. This holomophic curve is just the F-theory description of
the (p,q)-string, which we is looking for. In fact, the complex
variable $s$ is the local parameter of this holomophic curve.
 It is obvious from the Eqs.\ref{no12} and Eqs.\ref{no14} that
 this holomorphic curve is a 2-cycle of K3.
Such $2$-cycle is a fibration of a (p,q) $1$-cycle of the torus
over a line segment connecting two points associated to the
positions of two [p,q] $7$-branes on the base.

\subsection{F-theory description of 3-string junction}

It should be pointed out that though the Eqs.\ref{no18}-\ref{no20}
are given in terms of a local affine coordinates, these equations
can be generalized to the entire K3 by projectivizing the affine
coordinates. On the other hand, the Eqs.\ref{no18}-\ref{no20}
give, in fact, a generic F-theory description of any string
junction or string network by replacing the Eq.\ref{no15} with the
more general one. For example, we can easily obtain the F-theory
picture of any 3-string junction by lifting the junction into K3.
More precisely, we can use the equation (4.5) in Ref.[30] as a
substitute for the above equation 15, and use the new
corresponding functions $H$ and $P$ as a substitute for the old
$H$ and $P$ in the Eqs.\ref{no18}-\ref{no20}. In fact, the forms
of these embedding equations, which represent the holomophic curve
in the elliptic K3, are not changed except substituting the old
functions $H$ and $P$ with the new counterparts of them.

   Thus, the Eqs.\ref{no18}-\ref{no20} provide an
unifying F-theory description of both string and string junction,
and is advantageous over the M-theory picture of them. This also
shows that the string junction is on the same footing as the
various (p,q)-string though they seem to be very different at
first sight. Moreover, as these 2-cycles are holomorphic, the
$(p,q)$-string and the junction are BPS, thus stable.

\section{Discussion}

      A geometrical description of states in supersymmetric field and string theories is very important to the full understanding of the theories. In this paper, we give
 a general geometrical description of BPS string and string junction from the F-theory viewpoint.

As there is a transition between the string and string junction in
some regions of the moduli space of the 7-brane positions, thus,
the string junction is, in fact, on the same footing as the
various (p,q)-string though they seem to be very different at
first sight. The F-theory picture in this paper provide an
unifying description of both string and string junction. We can
easily see, in fact, a continuous transition between them because
they are on the same footing in Eqs.\ref{no18}-\ref{no20} except
the difference between the forms of the old functions $H$, $P$ and
the new ones, associated to the string and string junction
respectively. Thus this is a meaningful picture, which may, in
some sense, captures some interesting geometric features.

\large{\bf Acknowledgments}
\small

      I would like to thank Prof. Wu Ke and  other peoples in the Institute of Theoretical Physics of the Chinese Academy of Sciences for helps in the period of my stay there as a Post-doctor. I am grateful to Prof. Bo-Yuan Hou for helps for long time. I would like to thank my family for support.
      I would also like to thank  Prof. Nobuyoshi Ohto who let me know their papers [39] on Nov. 2003.

\end{document}